\documentclass[prd,floatfix,onecolumn,amsmath,amssymb,floatfix]{revtex4}

\usepackage[
pdfauthor={derajan},
pdftitle={How to do this},
pdfstartview=XYZ,
bookmarks=true,
colorlinks=true,
linkcolor=red,
urlcolor=blue,
citecolor=blue,
pdftex,
bookmarks=true,
linktocpage=true, 
hyperindex=true
]{hyperref}
\usepackage{hyperref}
\usepackage{natbib}
\usepackage[all]{hypcap}
\hypersetup{
backref=true
bookmarksnumbered,
pdfstartview={FitH},
citecolor={blue},
linkcolor={red},
urlcolor={purple!80!black},
pdfpagemode={UseOutlines}
}

\usepackage{graphicx,color,dcolumn,booktabs,bm}
\usepackage{subfigure}
\usepackage{longtable,lscape}
\usepackage{amssymb}
\usepackage{indentfirst}
\usepackage{epsfig}
\usepackage{feynmf}   
\usepackage{epstopdf}   
\usepackage{slashed}  
\usepackage{cases}
\usepackage[pdfpages]{xcolor}
\definecolor{maroon}{RGB}{139,25,150}
\usepackage{multirow}
\usepackage{float}
\usepackage{graphicx,color,dcolumn,booktabs,bm}

\begin{document}

\preprint{}
\preprint{}
\title{Properties of $P_c(4312)$ pentaquark and its bottom partner}

\author{K.~Azizi}
\email{ kazem.azizi@ut.ac.ir}
\affiliation{Department of Physics, University of Tehran, North Karegar Ave., Tehran
14395-547, Iran}
\affiliation{Department of Physics, Do\u gu\c s University,
Ac{\i}badem-Kad{\i}k\"oy, 34722 Istanbul, Turkey}
\author{Y.~Sarac}
\email{yasemin.sarac@atilim.edu.tr}
\affiliation{Electrical and Electronics Engineering Department,
Atilim University, 06836 Ankara, Turkey}
\author{H.~Sundu}
\email{ hayriye.sundu@kocaeli.edu.tr}
\affiliation{Department of Physics, Kocaeli University, 41380 Izmit, Turkey}

\date{\today}

\begin{abstract}
We present an analysis of the newly observed pentaquark $P_c(4312)^+$ to  shed light on its quantum numbers. To do that, the QCD sum rule approach is used. The measured mass of this particle is close to $\Sigma_c^{++}\bar{D}^-$ threshold and has a small width, which supports the  possibility of its being a molecular state. We consider an interpolating current in a molecular form and analyze both the positive and negative parity states with spin-$\frac{1}{2}$. We also consider the bottom counterpart of the state with similar molecular form. Our mass result for the charm pentaquark state supports that, the quantum numbers of the observed state are consistent with $J^P=\frac{1}{2}^{-}$.  
\end{abstract}
\maketitle

\vspace{-1mm}
\maketitle
\renewcommand{\thefootnote}{\#\arabic{footnote}}
\setcounter{footnote}{0}
\section{\label{sec:level1}Introduction}\label{intro}

The hadrons having nonconventional structures have been a subject of interest for many years. Since quantum chromodynamics (QCD) does not rule out their existence, they have been investigated extensively both theoretically and experimentally. Though their roots extend before the proposal of the  QCD, we had the first experimental evidence of such states in 2003 by the observation of the $X(3872)$~\cite{Choi:2003ue}. This observation has placed the exotic hadrons and their identification at the focus of interests. In the past decades, many exotic state candidates were observed~\cite{Choi:2003ue,Aubert:2006bu,Ablikim:2007ab,Liu:2013dau,Ablikim:2013mio,Ablikim:2013dyn,Ablikim:2013wzq,Ablikim:2014qwy,Ablikim:2015uix,BESIII:2016adj} and listed in the Review of Particle Physics (PDG)~\cite{PDG2018}. On the other hand, their internal structures are still not certain and there are lots of works devoted to the identification of their sub-structure. It seems that we will possibly witness more such exotic states in the future. Therefore, it is important to understand the nature and sub-structure of such observed states as well as provide information for the possible future observations via offering candidate states.    

The pentaquarks are among these exotic states, whose existence was controversial before 2015. In 2015 the LHCb collaboration investigated the $\Lambda_b\rightarrow J/\psi p K$ process and reported the observation of the two candidates for pentaquark states, namely $P_c(4380)^+$ and $P_c(4450)^+$~\cite{Aaij:2015tga}, in the $J/\psi p$ invariant mass distribution. After the observation of $P_c(4380)^+$ and $P_c(4450)^+$, different approaches were considered to clarify their inner structures and quantum numbers. They were interpreted as diquark-diquark-antiquark or diquark-triquark states~\cite{Lebed:2015tna,Li:2015gta,Maiani:2015vwa,Anisovich:2015cia,Wang:2015ava,Wang:2015epa,Wang:2015ixb,Ghosh:2015ksa,Zhu:2015bba,Wang:2016dzu} and meson baryon molecules~\cite{Chen:2015moa,Chen:2015loa,Meissner:2015mza,He:2015cea,Roca:2015dva,Azizi:2016dhy}. The observed peaks were discussed in Refs.~\cite{Guo1,Guo2,Mikhasenko:2015vca,Liu1} considering their possibility of being kinematical effect corresponding to triangle singularity as well. There were also investigations in the literature on the properties of other candidate pentaquark states with different quark contents such as strange hidden charm pentaquark states~\cite{Liu:2020cmw,Chen:2015sxa,Feijoo:2015kts,Lu:2016roh,Irie:2017qai,Chen:2016ryt,Zhang:2020cdi,Pimikov:2019dyr}, hidden bottom pentaquark states~\cite{Paryev:2020jkp,Gutsche:2019mkg,Azizi:2017bgs,Cao:2019gqo} and single or triple charmed pentaquark states~\cite{Azizi:2018dva}.

In the LHCb's recent analysis conducted by revisiting the $\Lambda_b^0\rightarrow J/\psi pK^-$ process with more accumulated data, the peak corresponding to $P_c(4450)^+$ state was observed to split into two peaks. Besides these two structures, labeled as $P_c(4440)^+$ and $P_c(4457)^+$, a new pentaquark state $P_c(4312)^+$ was also reported. The masses and widths for the observed resonances were reported as~\cite{Aaij:2019vzc}
\begin{eqnarray}
m_{P_c(4312)^+}&=&4311.9\pm 0.7^{+6.8}_{-0.6}~\mbox{MeV}~~~~~~~~\Gamma_{P_c(4312)^+}=9.8\pm 2.7^{+3.7}_{-4.5}~\mbox{MeV},\nonumber\\
m_{P_c(4440)^+}&=&4440.3\pm 1.3^{+4.1}_{-4.7}~\mbox{MeV}~~~~~~~~\Gamma_{P_c(4440)^+}=20.6\pm 4.9^{+8.7}_{-10.1}~\mbox{MeV},\nonumber\\
m_{P_c(4457)^+}&=&4457.3\pm 0.6^{+4.1}_{-1.7}~\mbox{MeV}~~~~~~~~\Gamma_{P_c(4457)^+}=6.4\pm 2.0^{+5.7}_{-1.9}~\mbox{MeV}.
\end{eqnarray}
This improvement has recollected the attentions toward these states. Complicated interactions between the participated quarks make it difficult to differentiate their internal structures. Although there are various possible interpretations about their substructure, their closeness to the meson baryon threshold and small width values are in favor of a molecular interpretation. Considering these the previously reported state $P_c(4450)^+$ was investigated taking into account the $\chi_{c1} p$ molecule~\cite{Meissner:2015mza}, $\bar{D}^*\Sigma_c$, $\bar{D}^* \Sigma_c^*$ molecule~\cite{Chen:2015loa,Chen:2015moa,Roca:2015dva,He:2015cea,Xiao:2015fia} or admixture of $\bar{D}\Sigma_c^*$ and $\bar{D}^*\Lambda_c$~\cite{Azizi:2016dhy} or $\bar{D}^*\Sigma_c$ and $\Lambda_c(2595)\bar{D}$ molecule~\cite{Geng:2017hxc,Burns:2015dwa}. However the recently released result, with analyses of more accumulated data, indicated that the peak corresponding to this state contains two separate ones. Therefore the suggested spin as $5/2$ may differ and these states may have a molecular form with smaller spins. Inspired by this motivation, following the new announcement of their observation by LHCb~\cite{Aaij:2019vzc}, the newly observed states, $P_c(4440)^+$, $P_c(4457)^+$ and $P_c(4312)^+$, were investigated via different methods. Based on their narrow widths and proximity of their masses to the $\bar{D}^*\Sigma_c$, $D\Sigma_c$, and $\bar{D}\Sigma_c$ thresholds, they were considered in various recent studies within the molecular interpretation. Among these studies that consider molecular sub-structure possibility is the one-boson-exchange model~\cite{Chen:2019asm,Wang:2019nwt}. In Ref.~\cite{Chen:2019asm} the $P_c(4312)^+$, $P_c(440)^+$, and $P_c(4457)^+$ were interpreted as loosely bound states  as $\Sigma_c\bar{D}$ with $J^P=\frac{1}{2}^-$ for $P_c(4312)^+$ and $\Sigma_c\bar{D}^*$ for $P_c(4440)^+$ and $P_c(4457)^+$ with $J^P=\frac{1}{2}^-$ and $\frac{3}{2}^-$, respectively. The molecular interpretation was also considered in the heavy hadron chiral perturbation theory~\cite{Meng:2019ilv}. With the quark delocalization current screening model~\cite{Liu:2019sip} pentaquark systems with quark content $uudd\bar{s}$ were considered in molecular interpretation. The QCD sum rules method was also used by adopting molecular pentaquark structure in Refs.~\cite{Chen:2019bip,Zhang:2019xtu,Wang:2018waa}. The Ref.~\cite{Chen:2019bip} interpreted the $P_c(4312)^+$ as $\Sigma_c^{++}\bar{D}^-$ bound state with $J^P=\frac{1}{2}^-$, and the $P_c(4440)^+$ and $P_c(4457)^+$ as $\Sigma_c^+\bar{D}^0$ with $J^P=\frac{1}{2}^-$, $\Sigma_c^{*++}\bar{D}^-$ and $\Sigma_c^+\bar{D}^{*0}$ with $J^P=\frac{3}{2}^-$ or $\Sigma_c^{*+}\bar{D}^{*0}$ with $J^P=\frac{5}{2}^-$. The $P_c(4312)^+$ state was considered in the Ref.~\cite{Zhang:2019xtu} as $\Sigma_c\bar{D}$ and its mass range was attained as $4.07\sim 4.97$~GeV. In the Ref.~\cite{Wang:2018waa} the molecular pentaquark state consideration led to the assignment of $\bar{D}\Sigma_c$ with $J^P=\frac{1}{2}^-$ for the $P_c(4312)^+$, and $\bar{D}^{*}\Sigma_c$ with $J^P=\frac{3}{2}^-$ or $\bar{D}^{*}\Sigma_c^{*}$ with $J^P=\frac{5}{2}^-$ for $P_c(4440/4457)^+$. Using the quasipotential Bethe-Salpeter equation approach~\cite{He:2019ify} the two structures, $P_c(4440)^+$ and 
$P_c(4457)^+$, were assigned as $\Sigma_c\bar{D}^*$ state with $J^P=\frac{1}{2}^-$ and $\frac{3}{2}^-$, and the $P_c(4312)^+$ was assigned as $\Sigma_c\bar{D}$ bound state with $J^P=\frac{1}{2}^-$. The spin parity quantum numbers were predicted as $J^P=\frac{1}{2}^-$, $J^P=\frac{3}{2}^-$  and $J^P=\frac{1}{2}^-$ for $P_c(4312)^+$, $P_c(4440)^+$ and $P_c(4457)^+$, respectively, in the quark delocalization color screening model~\cite{Huang:2019jlf}. Another molecular picture consideration gave their quantum numbers as $J^P=\frac{1}{2}^-$, $J^P=\frac{1}{2}^-$ and $J^P=\frac{3}{2}^-$ for $P_c(4312)^+$, $P_c(4440)^+$ and $P_c(4457)^+$, respectively~\cite{Xiao:2019aya}. They were also considered within the hadrocharmonium scenario~\cite{Eides:2019tgv} in which the  $P_c(4440)^+$ and $P_c(4457)^+$ were interpreted as almost degenerate states with respective spin parities $J^P=\frac{1}{2}^-$ and $J^P=\frac{3}{2}^-$ while the $P_c(4312)^+$ was interpreted as $J^P=\frac{1}{2}^+$. In their identification, the diquark-diquark-antiquark framework was applied with different approaches~\cite{Wang:2019got,Semenova:2019gzf,Ali:2019clg,Ali:2019npk}. In Ref.~\cite{Wang:2019got} the spins were stated as $J^P=\frac{1}{2}^-$ for $P_c(4312)^+$, $J^P=\frac{1}{2}^-$, $J^P=\frac{3}{2}^-$ or $J^P=\frac{5}{2}^-$ for $P_c(4440)^+$ and $J^P=\frac{1}{2}^-$ or $J^P=\frac{3}{2}^-$ for $P_c(4457)^+$ state through the predicted masses using the QCD sum rule method. Ref.~\cite{Semenova:2019gzf} considered the spin parity quantum numbers of $P_c(4312)^+$, $P_c(4440)^+$ and $P_c(4457)^+$ states as $J^P=\frac{1}{2}^-$, $J^P=\frac{1}{2}^-$ and $J^P=\frac{3}{2}^-$, respectively. The suggested spin parity quantum numbers in Refs.~\cite{Ali:2019clg,Ali:2019npk} are  $J^P=\frac{3}{2}^-$, $J^P=\frac{3}{2}^+$ and $J^P=\frac{5}{2}^+$ for $P_c(4312)^+$, $P_c(4440)^+$ and $P_c(4457)^+$, respectively. These states were also investigated through their production and decay mechanisms, see for instance the Refs.~\cite{Azizi:2018bdv,Wang:2019hyc,Wang:2020rdh,Xu:2019zme,Guo:2019fdo,Xiao:2019mvs}. 

On the other hand, one can not place aside the possible existence of b-partner states with quark sub-structures in which the charmed quark is replaced by the bottom one.  In history, generally, the observations of the states with charmed quarks were followed by identification of their bottom counterparts. Therefore, the expectation of a similar possibility for pentaquarks is quite natural. With this expectation, the experimental searches have also headed for their investigations, and for instance, the LHCb collaboration investigated the pentaquark states with a single bottom quark~\cite{Aaij:2017jgf}. This also indicates similar investigations may take place for hidden bottom pentaquarks in the future. This prospect also attracted theoretical searches, and various features of the pentaquarks with bottom quark were investigated~\cite{Paryev:2020jkp,Gutsche:2019mkg,Azizi:2017bgs,Cao:2019gqo,Shimizu:2016rrd,Lin:2018kcc,Peng:2019wys} to shed light on experimental seeking. Pursuing this aim, the possible hidden bottom pentaquark states with $J^P=\frac{1}{2}^\pm$, $\frac{3}{2}^\pm$, and $\frac{5}{2}^\pm$ were investigated within a chiral quark model~\cite{Yang:2018oqd}. In Ref.~\cite{Shen:2017ayv} a unitary coupled-channel model was used to study $B\Lambda_b-B\Sigma_b$ interactions, and the pole positions were predicted for the hidden bottom pentaquark states with different spin-parities. In Refs.~\cite{Huang:2018wed,Huang:2015uda} the masses and decay widths for the hidden charm and hidden bottom pentaquarks with different spin parities were predicted within the quark delocalization color screening model. The resonance mass range of $\Sigma_bB$ state was predicted as $11077.2 \sim 11079.8$~MeV~\cite{Huang:2018wed} and it was obtained as $11070$~MeV in Ref.~\cite{Huang:2015uda}.

As is seen, for the pentaquark states, identifying their structures and quantum numbers are among  the main issues. Despite many studies, there is no conclusive consensus on their sub-structure. Therefore it is important to investigate their structure deeply through different approaches. These studies are also helpful to check and better understand the results of different works.  In this work, we analyze the recently observed pentaquark state $P_c(4312)^+$ through the investigation of its mass via the QCD sum rules approach~\cite{Shifman:1978bx,Shifman:1978by,Ioffe81}.  Additionally, we consider the bottom counterpart of this state. The QCD sum rule method was widely used in the literature to gain information about the properties of hadrons giving successful predictions quite consistent with the experimental ones. To this end, we apply interpolating currents for these states in the molecular form and we consider both the negative and positive parity cases with spin-$\frac{1}{2}$ quantum numbers. The predictions attained for the masses are compared with the experimental observation to conjecture the possible spin parity quantum numbers of the $P_c(4312)^+$ state. We also obtain the current coupling constants, which can be adopted as input in searching the decay properties of the considered states. After the observation of $P_c(4380)^+$ and $P_c(4450)^+$~\cite{Aaij:2015tga}, in Ref.~\cite{Azizi:2016dhy}, these states were considered in a similar manner. Using molecular interpolating currents for $P_c(4380)^+$ and $P_c(4450)^+$ states with spin parity quantum numbers $J^{P}=\frac{3}{2}^\pm$ and $J^{P}=\frac{5}{2}^\pm$, respectively, they were investigated via the QCD sum rules method. The assigned possible spin-parity quantum numbers for $P_c(4450)^+$ state, $J^P=\frac{5}{2}^+$, with mass $m=4.44\pm 0.15$~GeV still shows consistency with the $P_c(4440)^+$ state after the report of the separation of $P_c(4450)^+$ into two peaks.   

The article has the following structure: Section~\ref{II} contains the details of the QCD sum rules calculation for the mentioned spectroscopic parameters. Section~\ref{III} gives the numerical analyses of the QCD sum rule results obtained analytically in the previous section.  The last section involves a summary and the comparison of the obtained results with experimental and other theoretical predictions.

\section{QCD sum rules calculations}\label{II}

To gain the mass and current coupling constant of the interested state via the QCD sum rule method, the calculation starts with the following two-point correlation function:
\begin{eqnarray}
\Pi(q)=i\int dx e^{iqx}\langle 0|\mathcal{T}\{J(x)\bar{J}(0)\}|0\rangle , \label{eq:CorrF1}
\end{eqnarray} 
where $\mathcal{T}$ denotes the time ordering operator. In the correlation function $J(x)$ represents the interpolating field for the state, and it is formed considering the valance quark content and quantum numbers of the state using quark fields. The related interpolating current in this analysis has the following form:
\begin{eqnarray}
J=\big[\epsilon^{abc}(u_a^T C\gamma_{\mu} u_b)\gamma^{\mu}\gamma_5c_c\big]\big[\bar{c}_d\gamma_5d_d\big],
\label{eq:interpolatingcurrent}
\end{eqnarray} 
with $u$, $d$, and $c$ being the respective quark fields, $a$, $b$ and $c$ subindices are the color indices and $C$ is the charge conjugation matrix.  For the analyses of the bottom counterpart of the $P_c$ pentaquark state, one can use the same interpolating current, $J$, given for $P_c$ pentaquark state in Eq.~(\ref{eq:interpolatingcurrent}) by the replacement of the $c$ quark fields with $b$ quark fields.

To calculate the correlation function, there are two approaches. In the first one, it is calculated in terms of hadronic degrees of freedom. The result from this side emerges in terms of the hadronic parameters such as mass and current coupling constant. Therefore this side is called as hadronic side or physical side of the calculation. The second approach in the calculation is the computation of the correlator in terms of the QCD degrees of freedom which gives the result in terms of QCD degrees of freedom, such as quark-gluon condensates, quark masses, QCD coupling constant. Therefore this part is called QCD or theoretical side. The QCD sum rules giving the physical quantities are obtained matching both sides of the calculations via dispersion relation. To improve this match providing suppression of the contributions of higher states and continuum, Borel transformation is applied to both sides. The results from both sides contain various Lorentz structures and the matching is performed considering the coefficient of the same Lorentz structure for both sides. 

For the calculation of the physical side, one treats the interpolating current as the operator annihilating or creating the hadron and inserts a complete set of the hadronic states in between the interpolating currents. One should note that the interpolating current couples both the negative and positive parity states. This application results in the following form of correlator after the computation of the integral over four$-x$
\begin{eqnarray}
\Pi^{\mathrm{Had}}(q)= \frac{\langle 0|J|P_{c(b)}(q,s):\frac{1}{2}^+\rangle \langle P_{c(b)}(q,s):\frac{1}{2}^+|\bar{J}|0\rangle}{m_{+}^2-q^2}+\frac{\langle 0|J|P_{c(b)}(q,s):\frac{1}{2}^-\rangle \langle P_{c(b)}(q,s):\frac{1}{2}^-|\bar{J}|0\rangle}{m_{-}^2-q^2}+\cdots,
\label{eq:hadronicside1}
\end{eqnarray}
where $\cdots$ represents the contribution of higher states and continuum. The matrix elements in the last results are represented in terms of the current coupling constants, $\lambda$, and Dirac spinor, $u(q,s)$, with spin $s$ as 
\begin{eqnarray}
\langle 0|J|P_{c(b)}(q,s):\frac{1}{2}^+\rangle &=& \lambda_{+} \gamma_5 u(q,s), \nonumber\\
\langle 0|J|P_{c(b)}(q,s):\frac{1}{2}^-\rangle &=& \lambda_{-} u(q,s).
\label{eq:matrixelement}
\end{eqnarray}
The $|P_{c(b)}(q,s):\frac{1}{2}^{+}\rangle$ represents one-particle $P_{c}$ $(P_{b})$ state with positive parity and mass $m_{+}$, and $|P_{c(b)}(q,s):\frac{1}{2}^{-}\rangle$ represents the state with opposite parity and mass $m_{-}$ and $\lambda_{+}$, $\lambda_{-}$ are their corresponding current coupling constants.  When  Eq.~(\ref{eq:matrixelement}) is used in  Eq.~(\ref{eq:hadronicside1}) and the summation over spins is applied using 
\begin{eqnarray}
\sum_s u(q,s)\bar{u}(q,s)=\not\!q+m,
\end{eqnarray}
the hadronic side becomes
\begin{eqnarray}
\Pi^{\mathrm{Had}}(q)=\frac{\lambda_{+}^2(\not\!q-m_{+})}{m_{+}^2-q^2}+\frac{\lambda_{-}^2(\not\!q+m_{-})}{m_{-}^2-q^2}+\cdots .
\end{eqnarray}
This result turns into 
\begin{eqnarray}
\tilde{\Pi}^{\mathrm{Had}}(q)=\lambda_{+}^2e^{-\frac{m_{+}^2}{M^2}}(\not\!q-m_{+})+\lambda_{-}^2e^{-\frac{m_{-}^2}{M^2}}(\not\!q+m_{-})+\cdots,
\end{eqnarray}
after the Borel transformation. The $\tilde{\Pi}^{\mathrm{Had}}(q)$ represents the Borel transformed form of the correlation function with $M^2$ being the Borel parameter. The result emerges with two Lorentz structures, namely $\not\!q$ and $I$. In the analyses, the coefficients of these structures for both hadronic and QCD sides are considered, and match of the coefficients of the same structures from both sides leads to the QCD sum rules for the considered physical quantities. Therefore the next step is the calculation of the coefficients of these structures from the QCD side.

The QCD side is computed using the interpolating current explicitly inside the correlation function, Eq.~(\ref{eq:CorrF1}). The quark fields are contracted via Wick's theorem and this application renders the result into the one given in terms of heavy and light quark propagators as
\begin{eqnarray}
\Pi^{\mathrm{QCD}}(q)&=&i\int d^4x e^{iqx} \epsilon_{abc}\epsilon_{a'b'c'}\Big[-\mathrm{Tr}[S_u(x)^{aa'}\gamma_{\mu'}CS_u^{T}{}^{bb'}(x)C\gamma_{\mu}]+\mathrm{Tr}[S_u(x)^{ba'}\gamma_{\mu'}CS_u^{T}{}^{ab'}(x)C\gamma_{\mu}]\Big]\nonumber\\
&\times & \mathrm{Tr}[\gamma_5S_d(x)^{dd'}\gamma_{5}S_{c(b)}^{d'd}(-x)]\gamma_{\mu}\gamma_5 S_{c(b)}^{cc'}(x)\gamma_5\gamma_{\mu'}.
\label{Eq:PiQCD}
\end{eqnarray}   
The propagators represented as $S_{u(d)}$ and $S_{c(b)}$ are light and heavy quark propagators with the following explicit forms:
\begin{eqnarray}
S_{q,}{}_{ab}(x)&=&i\delta _{ab}\frac{\slashed x}{2\pi ^{2}x^{4}}-\delta _{ab}%
\frac{m_{q}}{4\pi ^{2}x^{2}}-\delta _{ab}\frac{\langle \overline{q}q\rangle
}{12} +i\delta _{ab}\frac{\slashed xm_{q}\langle \overline{q}q\rangle }{48}%
-\delta _{ab}\frac{x^{2}}{192}\langle \overline{q}g_{\mathrm{s}}\sigma
Gq\rangle +i\delta _{ab}\frac{x^{2}\slashed xm_{q}}{1152}\langle \overline{q}%
g_{\mathrm{s}}\sigma Gq\rangle  \notag \\
&&-i\frac{g_{\mathrm{s}}G_{ab}^{\alpha \beta }}{32\pi ^{2}x^{2}}\left[ %
\slashed x{\sigma _{\alpha \beta }+\sigma _{\alpha \beta }}\slashed x\right]
-i\delta _{ab}\frac{x^{2}\slashed xg_{\mathrm{s}}^{2}\langle \overline{q}%
q\rangle ^{2}}{7776} ,  \label{Eq:qprop}
\end{eqnarray}%
and
\begin{eqnarray}
S_{Q,{ab}}(x)&=&\frac{i}{(2\pi)^4}\int d^4k e^{-ik \cdot x} \left\{
\frac{\delta_{ab}}{\!\not\!{k}-m_Q}
-\frac{g_sG^{\alpha\beta}_{ab}}{4}\frac{\sigma_{\alpha\beta}(\!\not\!{k}+m_Q)+
(\!\not\!{k}+m_Q)\sigma_{\alpha\beta}}{(k^2-m_Q^2)^2}\right.\nonumber\\
&&\left.+\frac{\pi^2}{3} \langle \frac{\alpha_sGG}{\pi}\rangle
\delta_{ij}m_Q \frac{k^2+m_Q\!\not\!{k}}{(k^2-m_Q^2)^4}
+\cdots\right\},
\end{eqnarray}
where $q$ is used for $u$ and $d$ quarks and $Q$ represents $c$ or $b$ quarks, $G^{\alpha\beta}_{ab}=G^{\alpha\beta}_{A}t_{ab}^{A}$ and $GG=G_{A}^{\alpha\beta}G_{A}^{\alpha\beta}$ with $a,~b=1,2,3$. Here $A=1,2,\cdots,8$ and $t^A=\frac{\lambda^A}{2}$ where $\lambda^A$ is the  Gell-Mann matrices. The propagators are used in  Eq.~(\ref{Eq:PiQCD}) explicitly and then Fourier and Borel transformations are applied. These steps result in 
\begin{eqnarray}
\tilde{\Pi}_i^{\mathrm{QCD}}(s_0,M^2)=\int_{(2m_Q+2m_u+m_d)^2}^{s_0}dse^{-\frac{s}{M^2}}\rho_i(s)+\Gamma_i(M^2),
\label{Eq:Cor:QCD}
\end{eqnarray}
where $\rho_i(s)$ is the spectral density obtained from the imaginary part of the result as $\rho(s)=\frac{1}{\pi}\mathrm{Im}\Pi^{\mathrm{QCD}}_i$ and sub-index $i$ represents either of the Lorentz structures, $\not\!q$ and $I$. Since the $\rho_i(s)$ and $\Gamma_i(M^2)$ are lengthy functions, we will not present their explicit forms here to avoid overwhelming mathematical representations and concentrate on the results obtained from their analyses, which will be presented in the next section. In Eq.~(\ref{Eq:Cor:QCD}), $s_0$ is the threshold parameter that arises after the application of the continuum subtraction using quark-hadron duality assumption, and $M^2$ is the Borel parameter. The match of the coefficients of the same Lorentz structures from hadronic and QCD sides leads us to the following relations:
\begin{eqnarray}
\lambda_{+}^2e^{-\frac{m_{+}^2}{M^2}}+\lambda_{-}^2e^{-\frac{m_{-}^2}{M^2}}&=&\tilde{\Pi}_{\not\!q}^{\mathrm{QCD}}(s_0,M^2),
\label{sumruleq}
\end{eqnarray}
and
\begin{eqnarray}
-\lambda_{+}^2m_{+}e^{-\frac{m_{+}^2}{M^2}}+\lambda_{-}^2m_{-}e^{-\frac{m_{-}^2}{M^2}}&=&\tilde{\Pi}_{I}^{\mathrm{QCD}}(s_0,M^2).
\label{sumruleI}
\end{eqnarray}
In the calculations, we take into account both the positive and negative  parity states. To obtain the related sum rules  Eqs.~(\ref{sumruleq}) and (\ref{sumruleI}) and their derivatives with respect to ($-\frac{1}{M^2}$) are considered and these four coupled equations are solved for the four unknown quantities, namely, $m_{+}$, $m_{-}$, $\lambda_{+}$ and $\lambda_{-}$. These last results are used for analyzing either  the  $P_c$ or $P_b$ pentaquarks.

Here, we should note that the similar current is used in Refs. \cite{Chen:2019bip,Chen:2016otp} to extract the mass of $P_c$ states. In the present study, against these works, we simultaneously include both the negative and positive parity states that couple to the selected  interpolating current as we also mentioned previously. We extract the spectroscopic properties of both parities with simultaneous solving of the sum rules obtained above. Then the obtained mass results are compared to the experimental value  to fix the quantum numbers of the observed state, $Pc(4312)$.  In the present study, we also extract the values of the current couplings for both parities, which may be used as main inputs in investigation of the electromagnetic, weak or strong decays of these states. We also obtain the parameters of b-partner states that may help experimental groups in searching for $P_b$ states in the experiment. In what follows we shall explain the numerical analyses of the results obtained from the QCD sum rule calculations.

\section{Numerical Analyses}\label{III} 

In this section the results obtained in Section~\ref{II} are applied to obtain the numerical values of the spectral properties of the candidate $P_c(4312)$  and its opposite parity state.  Besides, here we also present the analyses for the bottom counterparts of these states. The analytic results contain some input parameters and auxiliary ones as well. Some of these input parameters used in the calculations are gathered in Table~\ref{tab:Inputs}.
\begin{table}[h!]
\begin{tabular}{|c|c|}
\hline\hline
Parameters & Values \\ \hline\hline
$m_{c}$                                     & $1.27\pm 0.02~\mathrm{GeV}$ \cite{PDG2018}\\
$m_{b}$                                     & $4.18^{+0.03}_{-0.02}~\mathrm{GeV}$ \cite{PDG2018}\\
$m_{u}$                                     & $2.16^{+0.49}_{-0.26}~\mathrm{MeV}$ \cite{PDG2018}\\
$m_{d}$                                     & $4.67^{+0.48}_{-0.17}~\mathrm{MeV}$ \cite{PDG2018}\\
$\langle \bar{q}q \rangle (1\mbox{GeV})$    & $(-0.24\pm 0.01)^3$ $\mathrm{GeV}^3$ \cite{Belyaev:1982sa}  \\
$m_{0}^2 $                                  & $(0.8\pm0.1)$ $\mathrm{GeV}^2$ \cite{Belyaev:1982sa}\\
$\langle \frac{\alpha_s}{\pi} G^2 \rangle $ & $(0.012\pm0.004)$ $~\mathrm{GeV}^4 $\cite{Belyaev:1982cd}\\
\hline\hline
\end{tabular}%
\caption{Some input parameters used in the analyses.}
\label{tab:Inputs}
\end{table} 
 
In addition to these input parameters, we have two auxiliary parameters needed in the analyses. These are the Borel parameter, $M^2$, and threshold parameter, $s_0$. These two parameters are  determined considering the analyses of the sum rules imposing some standard criteria.  These criteria include the mild dependence of the results on these auxiliary parameters, the dominance of the contributions of the states in question on the higher ones and continuum, and convergence of the operator product expansion (OPE). The working intervals of the auxiliary parameters are the ones for which these criteria are satisfied. To deduce the $M^2$ the contribution of higher ordered terms in the QCD side should be sufficiently small and the contribution of the lower states should dominate that of the higher states. To determine the lower limit of the $M^2$, we take into account the region where the contribution of the higher dimensional term in OPE is less than $\sim 1~\%$.  For the upper limit of the $M^2$ we consider the ratio of the pole term to total one as $PC$, that is:
\begin{eqnarray}
PC(M^2)=\frac{\tilde{\Pi}_{i}^{\mathrm{QCD}}(s_0,M^2)}{\tilde{\Pi}_{i}^{\mathrm{QCD}}(\infty,M^2)}
\end{eqnarray}
and in our analysis, we require this ratio to be larger than $\sim 27~\%$. Sticking to these criteria we fix the interval for the Borel parameters as
\begin{eqnarray}
5.5~\mbox{GeV}^2\leq M^2\leq 6.5~\mbox{GeV}^2,
\end{eqnarray}
for $P_c$ states and
\begin{eqnarray}
11.5~\mbox{GeV}^2\leq M^2\leq 13.5~\mbox{GeV}^2,
\end{eqnarray}
for for $P_b$ states. For the  determination of the threshold parameter, we consider the relative weak dependence of the results on this parameter. Accordingly, its working intervals for the analyses are set as
\begin{eqnarray}
&20.5~\mbox{GeV}^2 \leq s_0 \leq 22.5~\mbox{GeV}^2 &
\end{eqnarray} 
for $P_c$ states and 
\begin{eqnarray}
&132.0~\mbox{GeV}^2 \leq s_0 \leq 136.0~\mbox{GeV}^2 &
\end{eqnarray} 
for $P_b$ states.

Having determined the auxiliary parameters, the masses and the current coupling constants can be determined using their intervals and the input parameters given in Table~\ref{tab:Inputs}. The obtained results are as follows:
\begin{eqnarray}
m_{-}=4322\pm 342~\mbox{MeV}~~~~~~\mbox{and}~~~~~~\lambda_{-}=(0.24\pm 0.09)\times 10^{-3}~\mbox{GeV}^5,
\end{eqnarray}
\begin{eqnarray}
m_{+}=4776\pm 380~\mbox{MeV}~~~~~~\mbox{and}~~~~~~\lambda_{+}=(0.38\pm 0.12)\times 10^{-3}~\mbox{GeV}^5,
\end{eqnarray}
for the pentaquark states containing charm quarks, and 
\begin{eqnarray}
m_{-}=11087\pm 73~\mbox{MeV}~~~~~~\mbox{and}~~~~~~\lambda_{-}=(0.11\pm 0.03)\times 10^{-3}~\mbox{GeV}^5,
\end{eqnarray}
\begin{eqnarray}
m_{+}=11105\pm 78~\mbox{MeV}~~~~~~\mbox{and}~~~~~~\lambda_{+}=(0.18\pm 0.05)\times 10^{-3}~\mbox{GeV}^5,
\end{eqnarray}
for the pentaquark states containing bottom quarks.

The results contain the errors arising from the uncertainty carried by the determinations of the auxiliary parameters and other input parameters used in the analyses.  The order of errors in b-channel is considerably small compared to the charmed case and the results in b-channel are more stable with respect to the changes in auxiliary parameters.  Our results for the parameters of both the parities in both the c and b channels may be checked via different approaches.

\section{Conclusions}\label{Conc} 

The recent report of the LHCb for the pentaquark states included a new pentaquark state $P_c(4312)^+$ and a split in the peak of previously observed $P_c(4450)^+$ which are named as $P_c(4440)^+$ and $P_c(4457)^+$. There are different interpretations of their sub-structure and quantum numbers, which need to be elucidated with further investigations. Among the different probable options, one possible structure is in favor of their having molecular form. By this motivation, this work is devoted to the analysis of the $P_c(4312)^+$ through the QCD sum rule method and predictions for its spectroscopic parameters. To this end, an interpolating current is chosen in the $\Sigma_c^{++}\bar{D}^-$ molecular form with negative parity. Besides, the opposite parity state is also studied.  

The ambiguity in their quantum numbers and substructures put these pentaquark states at the focus of interests. Therefore these states were investigated via alternative models. Table~\ref{tab:resulst-table} presents predictions for the masses and possible $J^P$ quantum numbers obtained by some of these models and the experimental observation as well. This table also contains our result attained in the present study for $P_c(4312)^+$ and the results of its possible bottom counterpart.
%
%
%
\begin{table}[h!]
\begin{tabular}{|c|c|c|c|c|c|c|c|c|c|}
\hline
Resonance           &    &      $J^P$        &  \begin{tabular}[c]{@{}c@{}} This study  \\ $(\mbox{MeV})$ \end{tabular}   & \begin{tabular}[c]{@{}c@{}}  \cite{Chen:2019bip} \\(\mbox{GeV}) \end{tabular}  & \begin{tabular}[c]{@{}c@{}} \cite{Huang:2019jlf,Huang:2018wed}\\(\mbox{MeV})\end{tabular} & \begin{tabular}[c]{@{}c@{}} \cite{Ali:2019npk}\\ (\mbox{MeV})\end{tabular} & \begin{tabular}[c]{@{}c@{}} \cite{Xiao:2019aya}\\(\mbox{MeV})\end{tabular} & \begin{tabular}[c]{@{}c@{}} \cite{Wang:2019got}\\(\mbox{GeV})\end{tabular} &  \begin{tabular}[c]{@{}c@{}} Experiment  \\ $(\mbox{MeV})$ \end{tabular} 
 \\ \hline\hline
\multirow{3}{*}{$P_c(4312)$} & \multirow{3}{*}{$m$}  & ~$\frac{1}{2}^-$&  $4322\pm 342$ & $4.33^{+0.17}_{-0.33}$        & $4306.7 \sim 4311.3$                      &               & $4306.4$              & \begin{tabular}[c]{@{}c@{}}$4.31 \pm 0.11$ \\ $4.34 \pm 0.14$\end{tabular}      &\multirow{3}{*}{ $4311.9\pm 0.7^{+6.8}_{-0.6}$}                \\ \cline{3-9}  
&  & ~$\frac{1}{2}^+$&  $4776\pm 380$     &         &         &    &           &     &  \\ \cline{3-9}  
                        &     &    ~$\frac{3}{2}^-$    &        &                         &  & $4240$   &          &    &                                                                                                                                                            \\ \hline
\hline
Resonance           &        &    $J^P$          &  \begin{tabular}[c]{@{}c@{}} This study  \\ $(\mbox{MeV})$ \end{tabular}      & \begin{tabular}[c]{@{}c@{}} \cite{Huang:2018wed}\\(\mbox{MeV})\end{tabular} & \begin{tabular}[c]{@{}c@{}} \cite{Huang:2015uda}\\(\mbox{MeV}) \end{tabular}& \begin{tabular}[c]{@{}c@{}} \cite{Yang:2018oqd} \\ (\mbox{MeV}) \end{tabular} \\   \cline{1-7}\cline{1-7}
\multirow{2}{*}{$P_b$} & \multirow{2}{*}{$m$}& $~\frac{1}{2}^-$  &   $11087\pm 73$   &   $11077.2 \sim 11079.8  $          & $11070$               & $11072-11074$                            \\ \cline{3-7} 
         &   &        $~\frac{1}{2}^+$   &   $11105\pm 78$        &           &    &    \\  \cline{1-7}                \end{tabular}
\caption{Mass and possible quantum numbers predicted for $P_c(4312)^+$ resonance and its bottom partner from different studies.}
\label{tab:resulst-table}
\end{table}
\begin{table}[]
\begin{tabular}{|c|c|c|c|c|c|c|c|c|}
\hline
      & \multicolumn{2}{c|}{This study} & \multicolumn{4}{c|}{\cite{Azizi:2016dhy}} \\ \hline
State & \multicolumn{2}{c|}{$P_c$}    & \multicolumn{2}{c|}{$P_c$}   & \multicolumn{2}{c|}{$P_c$}  \\ \hline
$J^P$               &$\frac{1}{2}^-$& $\frac{1}{2}^+$ &$\frac{3}{2}^-$ &$\frac{3}{2}^+$ & $\frac{5}{2}^-$  & $\frac{5}{2}^+$ \\ \hline
Mass~(\mbox{MeV} )  &$4322\pm 342 $ & $4776\pm 380$   & $4300\pm 100$  & $4240 \pm 160$ & $4200 \pm 150$   &  $4440 \pm 150$             \\ \hline
\end{tabular}
\caption{Mass predictions for hidden-charm pentaquark states with different quantum numbers  from this  study and Ref.~\cite{Azizi:2016dhy}.  }
\label{tab:tableIII}
\end{table}

It is seen from this table that the central value of our mass prediction,  $m_{-}=4322\pm 342$, obtained for the negative parity spin-$\frac{1}{2}$ state is consistent with the experimental result for $P_c(4312)^+$ and the predictions of  different theoretical studies for this particle, which suggests the possible spin parity of this particle being $J^P=\frac{1}{2}^-$. And also we present a mass prediction for the opposite parity state with the same spin as $m_{+}=4776\pm 380~\mbox{MeV}$. However, this mass value is higher than the reported masses of the $P_c(4312)^+$ and the other observed pentaquark states, as well.

As is seen from Table~\ref{tab:resulst-table}, we also considered the bottom counterpart of the $P_c(4312)$ state with a molecular form $\Sigma_b^{+}\bar{B}^0$ that is labeled as $P_b$ in the table. The mass predictions for the negative and positive parity cases are presented as  $m_{-}=11087\pm 73$~MeV and $m_{+}=11105\pm 78$~MeV, respectively. The prediction for the negative parity case is compared with the other predictions present in the literature, and it is seen to be consistent with their results within the errors, as is seen from the table.

In Ref.~\cite{Azizi:2016dhy}, we considered the $P_c(4380)^+$ and $P_c(4450)^+$ states in the molecular form with $J^P=\frac{3}{2}^{\pm}$ and $J^P=\frac{5}{2}^{\pm}$, respectively, and  made the mass calculations for both the  positive and negative parity cases.  The obtained results are shown in  Table~\ref{tab:tableIII} for comparison with the results of the cases $J^P=\frac{1}{2}^{\pm}$.  In the calculations made in Ref.~\cite{Azizi:2016dhy},  we considered a molecular current for $P_c(4380)^+$ and  a mixed interpolating current in the molecular form which is an admixture of $\bar{D}\Sigma_c^*$ and $\bar{D}^*\Lambda_c$ for $P_c(4450)^+$.  In light of the new experimental report indicating the split in the peak of  $P_c(4450)^+$ state, our final remark for the state  $P_c(4450)^+$ in our previous work in Ref.~\cite{Azizi:2016dhy} still supports one of these new states, the state $P_c(4440)$, to possibly have spin parity quantum numbers $J^P=\frac{5}{2}^+$.
 As is seen from Table~\ref{tab:tableIII}, compared with the experimental value, the state with  $J^P=\frac{5}{2}^+$  has a good mass consistency with the experimental result,  $m_{P_c(4440)^+}=4440.3\pm 1.3^{+4.1}_{-4.7}$~MeV~\cite{Aaij:2019vzc}, and supports the molecular structure for this state. 

The investigations of pentaquark states resulted in support of different possibilities for their sub-substructures leaving their structures still ambiguous. Therefore, to discriminate their sub-structure, we need further theoretical and experimental investigations. The results obtained for the masses and current coupling constants of both the negative and positive parity hidden-charmed/hidden-bottom  states in the present study may supply inputs for these further investigations.


\end{document}